\documentclass[%
reprint,
superscriptaddress,
frontmatterverbose, 
nofootinbib,
amsmath,amssymb,
floatfix,
]{revtex4-1}

\usepackage{graphicx}
\usepackage{amsmath}
\usepackage{amssymb}
\usepackage{dcolumn}
\usepackage{bm}
\usepackage{natbib}
\usepackage{epstopdf}
\usepackage{rotating}
\usepackage{color}

\newcommand{\avg}[1]{\langle #1 \rangle}

\newcommand{\kB}{\ensuremath{k_{\texttt{B}}}}
\newcommand{\Nh}{\ensuremath{N_{\texttt{h}}}}
\newcommand{\Nl}{\ensuremath{N_{\texttt{l}}}}
\newcommand{\Mtot}{\ensuremath{M_{\texttt{total}}}}
\newcommand{\Pcm}{\ensuremath{\bm{P}_{\texttt{cm}}}}

\newcommand{\Pcmx}{\ensuremath{P_{\texttt{cm},x}}}
\newcommand{\Pcmy}{\ensuremath{P_{\texttt{cm},y}}}
\newcommand{\Pcmz}{\ensuremath{P_{\texttt{cmz},z}}}

\newcommand{\Ekin}{\ensuremath{E^{\text{kin}}}}
\newcommand{\Ekinh}{\ensuremath{E^{\text{kin}}_{\texttt{h}}}}
\newcommand{\myeq}{\ensuremath{\!=\!}}
\newcommand{\tauLJ}{\ensuremath{\tau_{\text{LJ}}}}
\newcommand{\lfree}{\ensuremath{l_{\text{free}}}}

\newcommand{\hide}[1]{}
\newcommand{\equref}[1]{Eq.~(\ref{#1})}
\newcommand{\figref}[1]{Fig.~(\ref{#1})}

\newcommand{\pd}[2]{\frac{\partial #1}{\partial #2}}

\begin{document}

\preprint{APS/123-QED}

\title{Maintaning the equipartition theorem\\ in small heterogeneous molecular dynamics ensembles}

\author{Nima H. Siboni}
\email{hamidi@mpie.de}
\affiliation{Aachen Institute for Computational Engineering Sciences (AICES), RWTH-Aachen, Germany.}
\affiliation{Max-Planck-Institute f\"{u}r Eisenforschung GmbH, D\"usseldorf, Germany.}
\author{Dierk Raabe}
\email{d.raabe@mpie.de}
\affiliation{Max-Planck-Institute f\"{u}r Eisenforschung GmbH, D\"usseldorf, Germany.}
\author{Fathollah Varnik}%
\email{Corresponding author: fathollah.varnik@rub.de}
\affiliation{Interdisciplinary Centre for Advanced Materials Simulation (ICAMS), Ruhr-Universit\"{a}t Bochum, Germany.}
\affiliation{Max-Planck-Institute f\"{u}r Eisenforschung GmbH, D\"usseldorf, Germany.}




\date{\today}

\begin{abstract}
It has been reported recently that the equipartition theorem is violated in molecular dynamics simulations with periodic boundary condition [Shirts et al, J. Chem. Phys. {\bf 125} 164102 (2006)]. This effect is associated with the conservation of the center of mass momentum. Here, we  propose a fluctuating center of mass molecular dynamics approach (FCMMD) to solve this problem. Using the analogy to a system exchanging momentum with its surroundings, we work out --and validate via simulations-- an expression for the rate at which fluctuations shall be added to the system. The restoration of equipartition within the FCMMD is then shown both at equilibrium as well as beyond equilibrium in the linear response regime.%
%
\end{abstract}

\maketitle

\section{\label{introduction}Introduction}
The equipartition theorem states that the total kinetic energy of a classical system in canonical ensemble is equally distributed among all degrees of freedom and that the average kinetic energy associated with the translational motion of a particle is given by $\avg{p^2/(2m)}=d \, \kB T/2$. Here, $p$ and $m$ are the momentum and mass of the particle. $\kB$ is the Boltzmann factor, $T$ denotes the temperature and $d$ is the spatial dimension. This relation serves to control the  temperature in molecular dynamics (MD) simulations by adjusting the kinetic energy of the system \cite{Bussi_JCP_2007}. 

It has been shown recently that in MD simulations with the periodic boundary condition (PBC) the equipartition theorem is violated \cite{Shirts_JCP_2006}. This effect is attributed to the conservation of the center of mass (or, equivalently, total) momentum, $\Pcm$, due to the PBC. This additional constant of motion restricts the simulation trajectories to only a subset of the phase space and leads to a difference between the time- and ensemble-averages \cite{Ray_PRE_1999,Tuckerman_JCP_2001}. It can be shown that, in canonical ensemble MD simulations with PBC, the average kinetic energy of a particle of mass $m$ obeys \cite{Uline_JCP_2008}
\begin{align}
\avg{\Ekin}=\frac{\avg{p^2}}{2m}=\frac{d\kB T}{2}\left(1-\frac{m}{\Mtot}\right),
\label{eq:pi2}
\end{align}
 where $\avg{\cdots}$ stands for statistical average and $\Mtot=\sum_{i=1}^N m_i$ is the total mass of the system ($N$ is the total number of particles). \equref{eq:pi2} shows that the violation of the equipartition theorem can be safely ignored if the mass of a single particle is negligible compared to the total mass of the system. Also, if all the particles have the same mass, $m/\Mtot=1/N$  the effect is negligible for most simulation cases. This is also in line with the general observation that differences  between to \textit{molecular dynamics ensembles}~\cite{Erpenbeck1977} and other thermodynamic ensembles become less significant as the system size grows and eventually vanishes in the thermodynamic limit ($N\to \infty$) \cite{Lebowitz_PR_1967}.

However, beyond iso-particle systems, the same problem arises principally also in hetero-particle systems, where for instace a small number of massive particles are surrounded by a large number of light particles. This includes explicit solvent MD simulations of transport properties of colloids and  nanoparticles in the dilute limit and generally all multi-component mixtures. In these cases, the violation of equipartitioning may not always be tolerable. For an estimate, let us consider a system consisting of $\Nl$ light particles of mass $m_0$ and $\Nh$ massive particles of mass $\alpha m_0$. This yields $\Mtot=(\Nl+\alpha\Nh)m_0$ and thus one obtains for the average kinetic energy of the heavy particle, $\avg{\Ekinh}=[1-\alpha/(\Nl + \alpha \Nh)]d\kB T/2$. Obviously, in the limit that the total mass of the heavy particles is large compared to the total mass of the light particles ($\alpha \gg \Nl/\Nh$), this reduces to $\avg{\Ekinh} \approx (1-1/\Nh)d \kB T /2$. Thus, the average kinetic energy of a single heavy particle approaches zero with increasing mass.

The present paper addresses this point. We propose a method to restore the equipartition theorem in molecular dynamics ensembles containating components with different masses. For this purpose, we first use computer simulations of a massive tracer particle in an ambient liquid and provide evidence for the idea that, as noted in \cite{Shirts_JCP_2006,Uline_JCP_2008}, the main cause of the problem is the conservation of the center of mass momentum. Based on this understanding, we propose a molecular dynamics method which allows for the fluctuations of the center of mass momentum, $\Pcm$ (note that, in equilibrium, $\delta  \Pcm=\Pcm$, since $\left<\Pcm\right>=0$). For a reliable implementation of the method, we also determine the rate at which fluctuations shall be added to the system, \equref{tau}. The validity of this expression is confirmed via computer simulations of  systems with rigid walls, which do not have the artificial constant motion, $\Pcm=0$. Finally, we test the method showing that it does restore equipartitioning both in equilibrium and beyond equilibrium in the linear response regime.

\section{Violation of equipartition for a massive tracer}

As mentioned above, a strong violation of the equipartition theorem is expected for the case of a massive tracer particle in a liquid environment. In order to demonstrate this property, we  perform MD simulations of a generic 80:20 binary mixture of Lennard-Jones particles (types  A and B) \cite{Kob1994,Kob1995}. A and B particles interact via
$U_{\text{LJ}}(r)\myeq
4\epsilon_{\alpha\beta}[(d_{\alpha\beta}/r)^{12}-(d_{\alpha\beta}/r)^6]$,
with $\alpha,\beta\myeq {\text{A,B}}$, $\epsilon_{\text{AB}}\myeq
1.5\epsilon_{\text{AA}}$, $\epsilon_{\text{BB}}\myeq
0.5\epsilon_{\text{AA}}$, $d_{\text{AB}}\myeq 0.8d_{\text{AA}}$,
$d_{\text{BB}}\myeq 0.88d_{\text{AA}}$, and $m_{\text{A}}\myeq
m_{\text{B}}$. The potential is truncated at twice the minimum
position of the LJ potential, $r_{\text{c},\alpha\beta} \myeq 2.245 d_{\alpha\beta}$. The parameters
$\epsilon_{\text{AA}}$, $d_{\text{AA}}$ and $m_{\text{A}}$ define
the units of energy, length and mass. The unit of time is given by $\tauLJ \myeq d_{\text{AA}}\sqrt{m_{\text{A}} / \epsilon_{\text{AA}}}$. The system density is kept constant at the value of $1.2$ and temperature at $T=1$ for all simulations whose results are reported here. Depending on the case studied, linear dimension and total particle number are in the range of $L \in [4.7, 203]$ and $N\in [125,10684]$. Equations of motion are integrated using the velocity-Verlet algorithm with a discrete time step of $dt \myeq 0.005$. 

The results presented here are expected to be largely model independent and hence general. The choice of the above model is purely historical and is motivated by the fact that we have been using it to study a number of problems in the context of the physics of glasses \cite{Varnik2004,Varnik2006b,Varnik2006d,Varnik2008b}. We indeed encountered the present problem of the violation of the equipartition theorem as we inserted massive tracer particles into our model to study the concept of effective temperature \cite{Sollich1998,Berthier2002a}.

With the exception of one, the mass of all particles is set to unity. One of the particles (of type B) is taken to be the massive tracer. The mass of this particle is then varied and its kinetic energy is monitored. Simulation results are averaged over 40 independent runs. In order to investigate the possible role of the thermostat, all the simulations are performed both for the Nos\'e-Hoover (N-H) \cite{Nose1984,Hoover1985} and the Andersen \cite{Andersen1980} thermostats. In equilibrium simulations, all components of particle velocities are coupled to the thermostat. We also extend the present analysis to a  non-equilibrium steady state situation by imposing a linear shear flow via the  SLLOD-algorithm combined with the Lees-Edwards boundary condition (LEBC) \cite{Evans1990}. In this case, coupling to the thermostat is done only for the velocity component in the direction perpendicular to the shear plane (vorticity direction). By doing so, we avoid problems related to the flow-induced bias on the kinetic energy, when regulating the system temperature \cite{Evans1986b}.

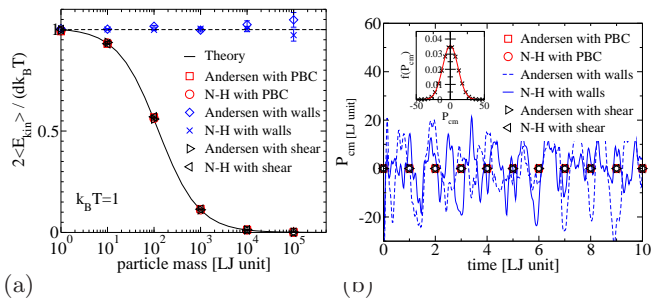
\begin{figure}
\unitlength=1mm
\begin{picture}(0,0)
\put(0,-2.5){(a)}
\put(45,-2.5){(b)}
\end{picture}
\includegraphics[width=4.2cm]{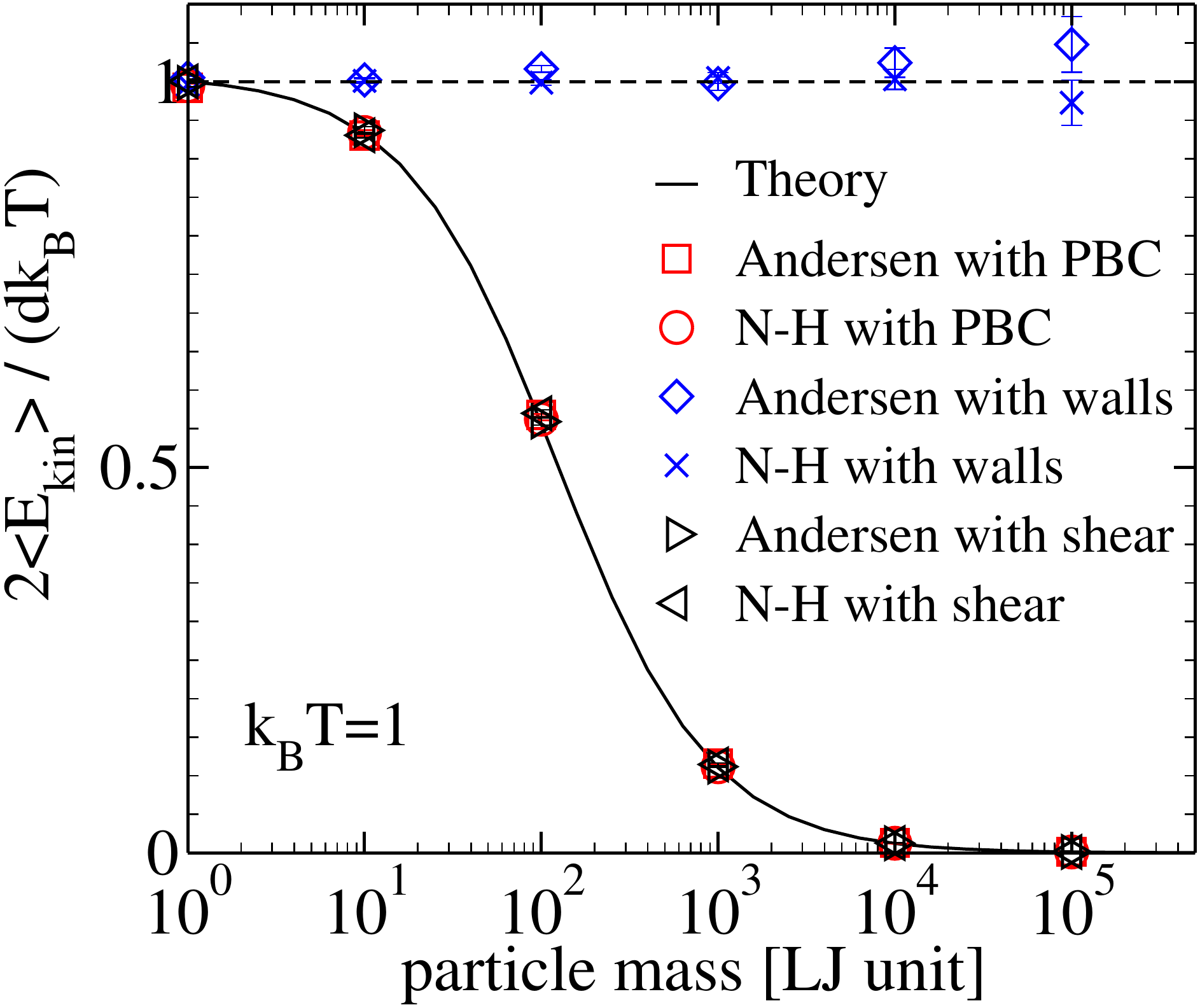}
\includegraphics[width=4.2cm]{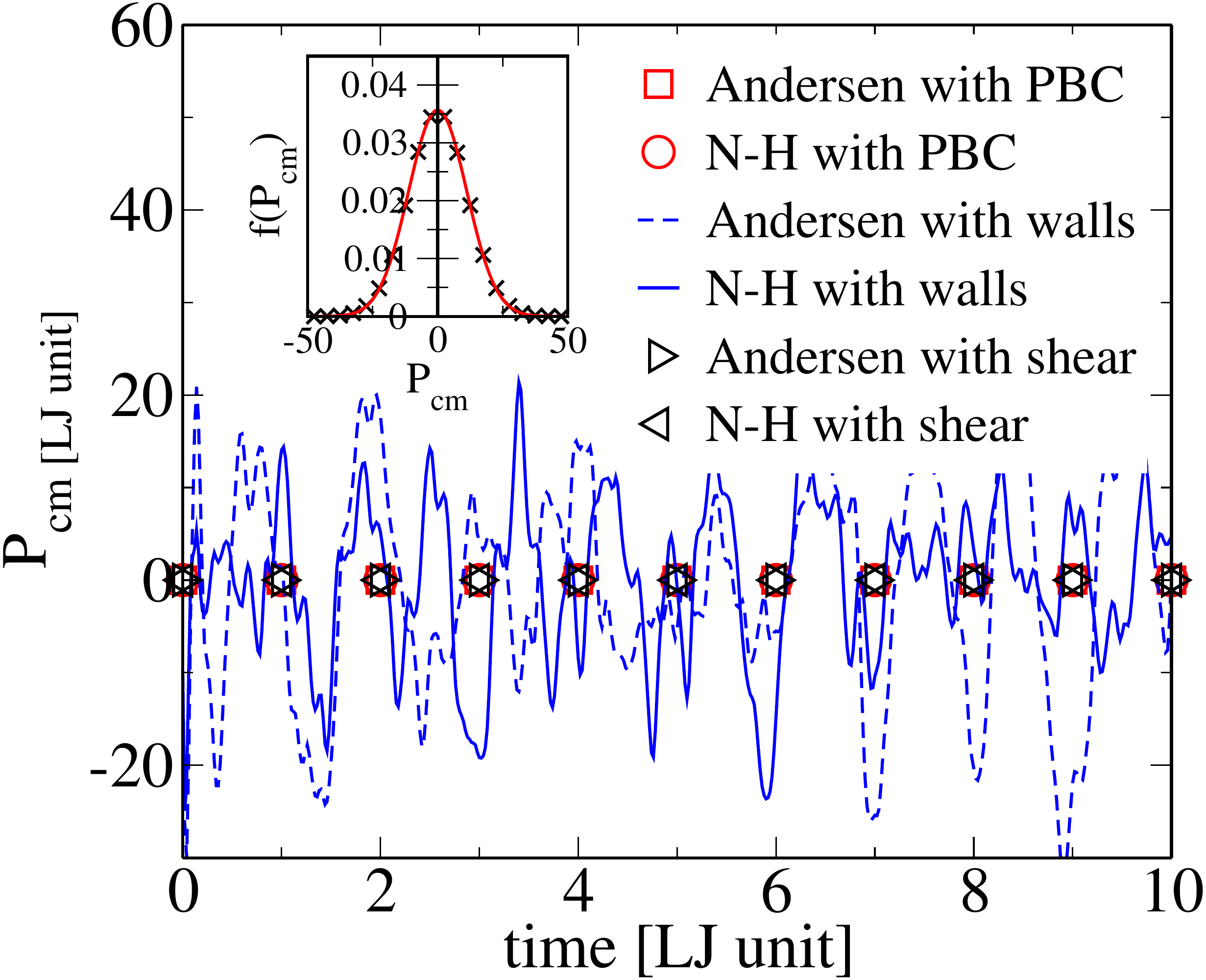}
\caption{(a) The average kinetic energy of a massive particle, normalized by $d\kB T/2$, is shown versus its mass for two popular thermostating methods, both in equilibrium and under steady shear (with uniform shear rate of $\dot{\gamma}=10^{-3}$). The solid line gives the analytic prediction given by \equref{eq:pi2}. Results for a system containing two planar walls are also shown for equilibrium simulations. While equipartition is violated in the simulations with PBC, it is well satisfied in the simulations with walls.  (b) Total momentum versus time for all the simulations reported in the panel (a). The inset shows $\Pcm$-distribution for the case of simulation with walls. It perfectly obeys the expected Gaussian distribution with zero mean and variance $\Mtot \kB T$ (The black dots are simulation results for $\Mtot \myeq 125$ and the continues line is the corresponding theoretical expectation). }
\label{fig:violation-of-equipartition}
\end{figure}

Results obtained via these simulations are depicted in \figref{fig:violation-of-equipartition}. As seen in this plot, the violation of the equipartition theorem occurs in perfect agreement with the theoretical predictions of \equref{eq:pi2}, independent of the specific thermostat.

Next we provide evidence from simulation that the deep reason for the violation of the equipartition theorem is indeed the conservation of the total momentum \cite{Shirts_JCP_2006,Uline_JCP_2008}. For this purpose, we have designed a simulation setup where $\Pcm$ is not conserved. This is achieved by introducing two  planar walls separated by a distance $L_z$ along the $z$-direction, while PBC is used along the $x$ and $y$ directions. The walls are made of particles with the same size and structure as the liquid particles so that liquid-wall interactions induce fluctuations of $\Pcm$ along all spatial directions. Results of these simulations are also shown in \figref{fig:violation-of-equipartition}, demonstrating that, as expected, the equipartition theorem is valid in systems with walls.

\section{Fluctuating center of mass MD (FCMMD)}
The above results suggest that a possibility to restore the equipartition theorem is to introduce walls with roughness on the particle scale. However, in studies focusing on bulk properties, walls are undesired since they in general influence the system properties unless very large wall-to-wall separations are used (see, e.g., \cite{Varnik2002c,Varnik2002e,Baschnagel2005} and references therein). Thus, it is desirable to introduce a method which uses PBC, while at the same time allowing for fluctuations of the total momentum. Such a method is proposed here. Our approach is quite simple and is motivated by the fact that, in a system exchanging momentum with its environment, the center of mass momentum is a fluctuating quantity.

Motivated by this idea, we perform the following two steps: (i) Draw a value for $\Pcm$ and (ii) distribute it among particles. These steps are carried over, repeatedly, during the simulation. In order to have the canonical sampling of the phase space, the total momentum assigned in step (i) should assume a distribution probability coinciding with the canonical distribution function for $\Pcm$:
\begin{align}\nonumber
\mathfrak{f}(\Pcm) &= \int f(\chi)\delta(\sum_{i=1}^N \bm{p}_i-\Pcm)d\chi\\\label{Pdist}
&= \sqrt{\frac{1}{2\pi \kB T \Mtot}}\exp\left[-\frac{\Pcm^2}{2 \kB T \Mtot}\right],
\end{align}
where $f(\chi)$ is the probability of the micro-state $\chi$, and $\delta$ is the Dirac delta-function. Similar to the discussion in Ref.~\cite{Bussi_JCP_2007} for choosing kinetic energy from the canonical distribution, one has a certain flexibility in choosing the sampling rate. Here, we provide a physical criterion to estimate the time scale of the $\Pcm$-fluctuations. We build our analysis upon the fact that fluctuations of $\Pcm$ are caused by the exchange of momentum with the surrounding medium.

For a system with $\Pcm(t=0)=0$, it follows from the above considerations that, due to interactions with the surrounding medium,  $\Pcm$ will not remain zero but increases with time. On the other hand, too large a value of $\Pcm$ will decay due to the same interactions. Collisions with the surrounding medium thus provide a source of stochastic noise and, at the same time, give rise to viscous friction. This is very similar to the fluctuations of the velocity of a Brownian particle in a fluid. The probability distribution of these fluctuations is obtained as the solution of a Fokker-Planck equation subjected to the potential $\phi$ \cite{Risken1989},
\begin{align}
\pd{\mathfrak{f}(\Pcm, t)}{t}
&=\mu \nabla|_{\Pcm} \cdot \left( \mathfrak{f} \nabla|_{\Pcm} \phi \right) +D \left(\nabla|_{\Pcm}\right)^2 \mathfrak{f},
\label{eq:FP}
\end{align}
where $\phi(\Pcm)=-\Pcm^2/(2\Mtot)$ and $\nabla|_{\Pcm}=(\partial / \partial \Pcmx, \partial / \partial \Pcmy, \partial / \partial \Pcmz)$. The mobility, $\mu$, and diffusion constant, $D$, obey the Einstein relation, $\mu=D/(\kB T)$. Given $\Pcm'$ at time $t'$, the conditional probability distribution at a time $t>t'$ is \cite{Risken1989},
\begin{align}\nonumber
\mathfrak{f}(\Pcm,t|\Pcm',t') = &\sqrt{\frac{1}{2\pi \sigma^2(t,t')}}\\
&\times \exp\left(  \frac{(\Pcm -\Pcm' \exp[-\gamma (t-t')])^2}{2 \sigma^2(t,t')}\right),
\label{eq:FP-solution}
\end{align}
where $\gamma=\mu/\Mtot$ and $\sigma^2(t,t')= d \kB T \Mtot [1-\exp(-2\gamma(t-t'))]$. It is seen from \equref{eq:FP-solution} that $\rho$ reaches the expected Maxwell distribution, \equref{Pdist}, in the limit of long times, $t-t' \gg 1/\gamma$. The characteristic time for reaching the equilibrium distribution of center of mass fluctuations is thus obtained from $\tau=1/\gamma$,
\begin{align}\label{chartime-a}
\tau=\frac{\Mtot \kB T}{D}.
\end{align}
This expression is not fully satisfactory as it contains an important unknown parameter, $D$. We therefore attempt at an estimate of $\tau$ from a microscopic consideration. For this purpose, we again recall that, starting with  $\Pcm(t=0)=0$ collisions with the surrounding medium will lead to $\left< \Pcm^2 \right>=  d \Mtot \kB T$ within a time of the order of $\tau$. For simplicity, we assume here that $\Pcm$ is the sum of $N_s$ statistically independent elementary momentum fluctuations, $\delta \bm{p}_i$, resulting from the collisions between fluid particles with the system's boundary, $\Pcm=\sum_i^{N_s} \delta \bm{p}_i$. This yields $\left<\Pcm^2 \right> = N_s \left<\delta p^2\right>$ and thus $N_s \left<\delta p^2\right>=d \Mtot \kB T$. The time scale $\tau$ is encoded in the number of elementary collisions $N_s$. To see this, we first note that momentum exchange occurs within a ``skin'' -- which runs parallel to the boundary -- 
of thickness equal to mean free path, $\lfree$. On average, $1/6^{\text{th}}$ of these particles in the skin layer move along the perpendicular direction toward the boundary and will undergo a collision within a time of $\delta t \sim \lfree / \avg{|v_\perp|} = \lfree / \sqrt{\kB T/m}$ where $\avg{|v_\perp|}$ is the average thermal velocity in the direction normal to the boundary. The total number of collisions within a time of  $\tau$ is thus obtained as $N_s \sim \rho \lfree A /6  \times \tau / \delta t \sim \rho A \tau \sqrt{\kB T/m}/6$. To arrive at a closed expression for $\tau$, the magnitude of the typical momentum exchange per collision is estimated: $\delta p \sim 2 m \avg{|v_\perp|}= 2 \sqrt{m\kB T}$. Combining the above two expressions for $N_s$ and using this last relation for $\delta p$, one finally finds
\begin{align}\label{tau}
\tau &=\frac{C}{\rho (m\kB T)^{1/2}}\frac{\Mtot}{A},
\end{align}
where $C$ is a constant prefactor. Equation (\ref{tau}) gives an estimate for the characteristic time of the $\Pcm$-fluctuations in a system exchanging momentum with its surroundings through a boundary (interface) of surface area $A$. 

\begin{figure}
\includegraphics[width=7cm]{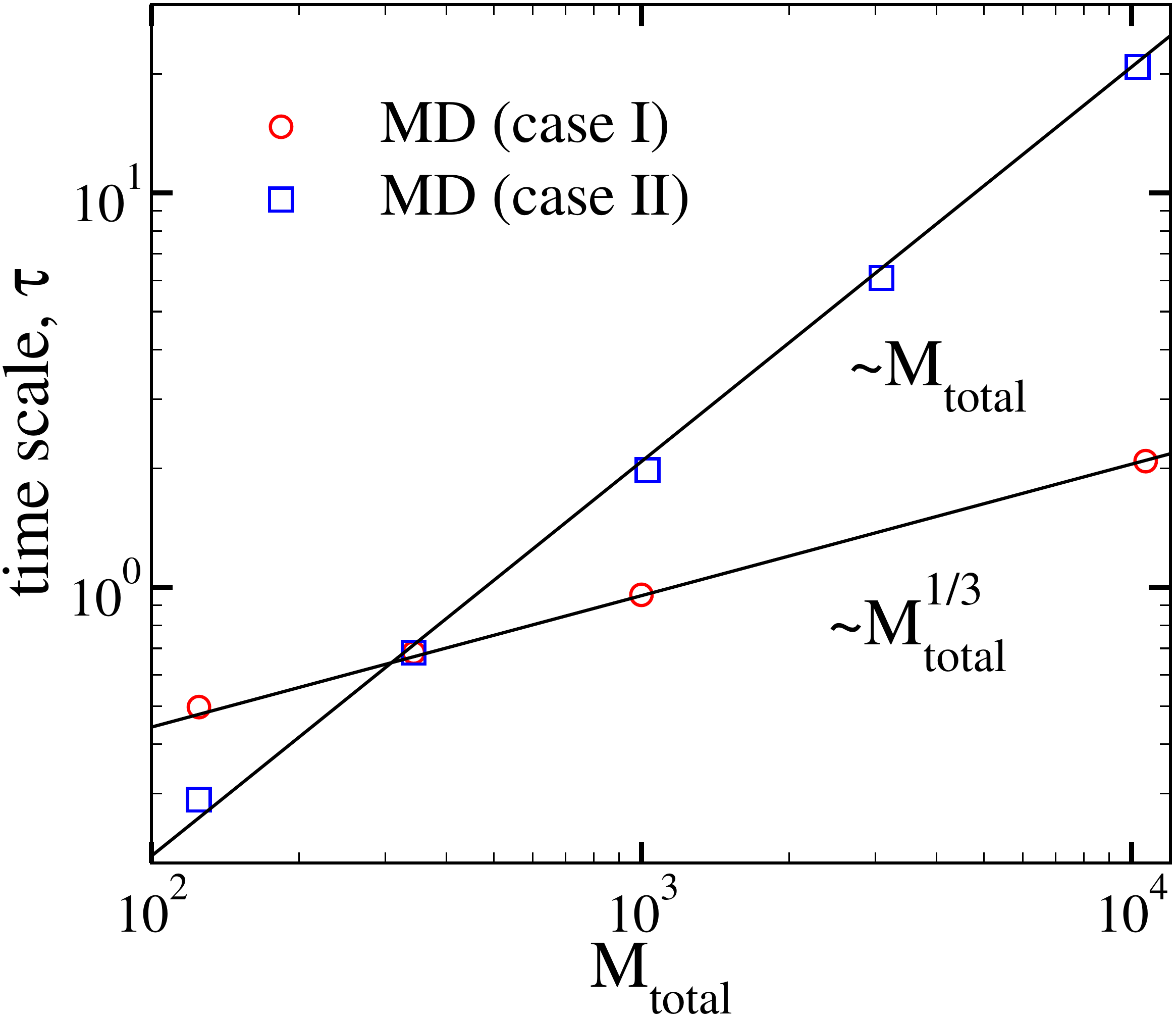}
\caption[]{The time scale, $\tau$, for $\Pcm$-fluctuations versus total mass in a system confined by two planar walls. $\tau$ is determined from the decay of the autocorrelation function, $\avg{\Pcm(t)\Pcm(0)}$. Results are shown for two different geometries. In case I, the simulation box is a cube and the variation of $\Mtot$ is accompanied by a corresponding change of the surface area of the walls. In case II, the surface area of the walls is kept constant but only the distance of the walls is varied. Using \equref{tau}, one thus expects $\tau \propto \Mtot^{1/3}$ in case I but $\tau \propto \Mtot$ in case II (solid lines).}
\label{fig:tau}
\end{figure}

In order to test this result, we have performed a series of three dimensional MD simulations of the present binary LJ model confined between two parallel walls for different system sizes while keeping all other simulation parameters constant (e.g., $T=1$, $\rho=1.2$, $m=1$). The characteristic time is measured by the auto-correlation time of fluctuations of $\Pcm$. Note that, in these simulations, PBC is used along the $x$ and $y$ directions, so that no momentum fluctuations will originate from the corresponding boundaries. In other words, the relevant surface area, $A$, appearing in \equref{tau} corresponds to the surface area of the walls. To better highlight the dependence of $\tau$ on $\Mtot$ and $A$, we studied two different geometries leading to qualitatively different results for $\tau$ in terms of the total mass. In the first series of simulations, the system was a cube with length $L$ so that $\Mtot=\rho L^3$ and $A=L^2=(\Mtot/\rho)^{2/3}$ and thus $\tau \propto \Mtot^{1/3}$ (case I). In the second series of simulations, we only varied the wall-to-wall separation, $L_z$, while keeping the surface area of the walls constant. This gives $\tau \propto \Mtot$ (case II). As shown in \figref{fig:tau}, results on the characteristic time of momentum exchange obtained for these two sets of simulations clearly confirm the validity of \equref{tau} with a constant of proportionality of $C \sim {\cal{O}}(1)$.

\begin{figure}
\includegraphics[width=7cm]{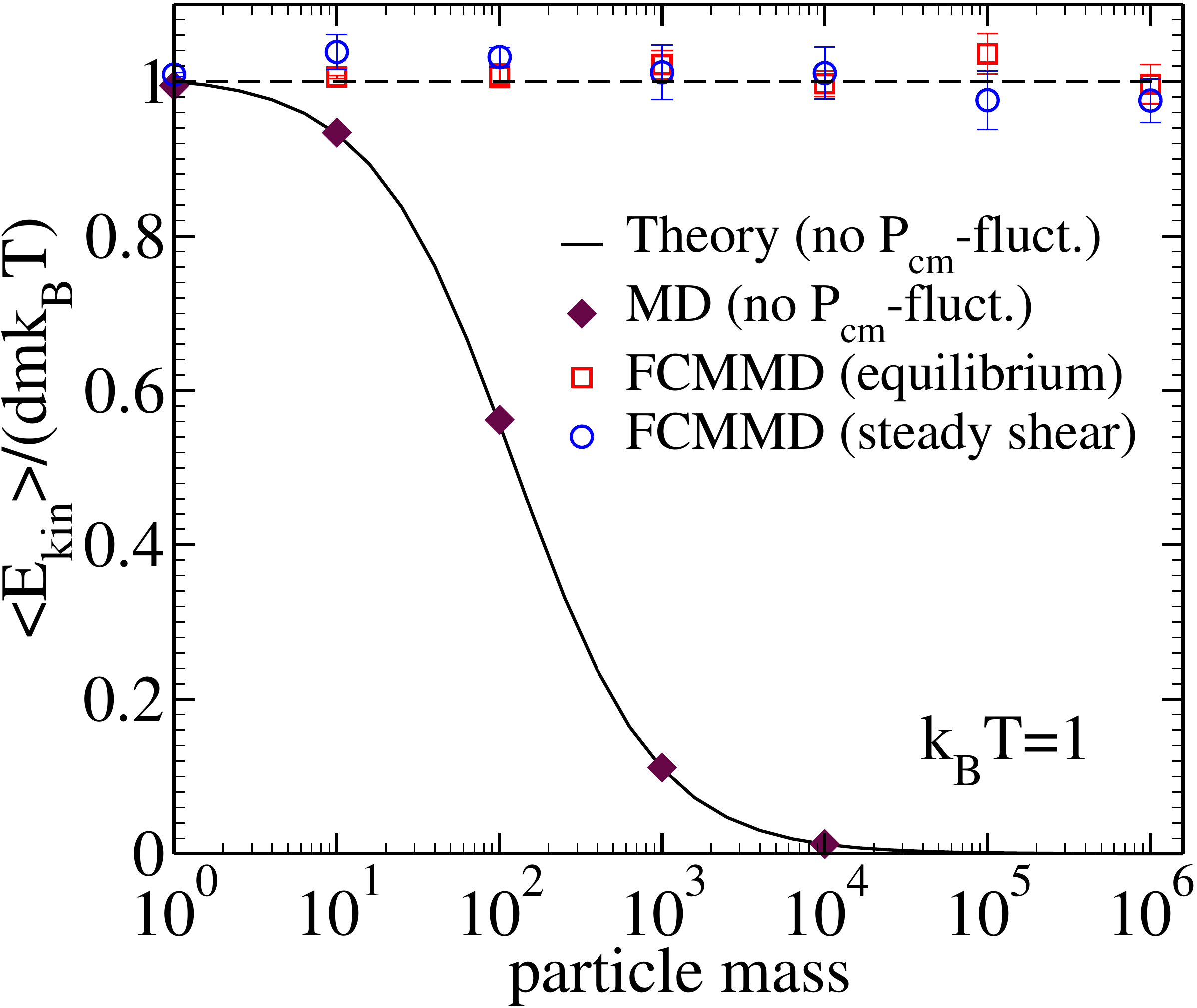}
\caption[]{Restoring the equipartition theorem via the proposed fluctuating center of mass molecular dynamics (FCMMD) method. The average kinetic energy of a massive particle, normalized by $d\,\kB T/2$, is shown versus its mass both for equilibrium simulations as well as under steady shear in the linear response regime (uniform shear rate of $\dot{\gamma}=10^{-3}$). For comparison, we also plot results of MD simulations without $\Pcm$-fluctuations and the corresponding theoretical curve, \equref{eq:pi2}.}
\label{fig:restoring-equipartition}
\end{figure}

In our scheme, we update $\Pcm$ using a random walk sampling $\mathfrak{f}(\Pcm)$ with time scale of order $\tau$, \equref{tau}. The question arises now as how to distribute a given $\Pcm$ among particles. In canonical ensemble, if the  total momentum of the system is $\Pcm$, the conditional average for the momentum of a particle is equal to $ \avg{p}|_{\Pcm}=m\Pcm/\Mtot$. We, therefore, propose that the imposed momentum change be divided among particles proportional to their individual masses. In analogy to the system with walls, the momentum change is applied to each particle once at a time. The order of particles is chosen randomly. After adding momentum to each particle, relative velocities of all particles with respect to the center of mass are rescaled. This last operation does not modify the center of mass momentum but allows to restore the kinetic energy exactly to the value before updating $\Pcm$.

Results obtained from these simulations are shown in \figref{fig:restoring-equipartition}. As shown in this figure, the proposed approach is able to restore the equipartition theorem both in equilibrium simulations as well as in a system beyond equilibrium in the linear response regime.

\section{Conclusion}\label{conclusion}

In this work, we propose a modification of the molecular dynamics method with periodic boundary condition to restore the equipartition theorem. The method is based on introducing fluctuations of the center of mass momentum. The issue of a proper rate at which $\Pcm$ fluctuations are imposed to the system is also addressed and validated against simulations. It is shown that the method restores equipartition both at equilibrium and under steady shear in the linear response regime. This latter finding is of crucial importance for studies, which focus on a violation of the equipartition due to non-linear off-equilibrium effects \cite{Sollich1998,Berthier2002a}. It is noteworthy that the violation of equipartition does not exclusively occur in MD simulations. As an example, it has also been observed in the fluctuating lattice Boltzmann method where equipartitioning is important at all length scales \cite{Ollila_JCP_2011}. The relevance of the present work is thus not restricted to MD simulations but may also provide guidance for restoring equipartitioning and hence, a correct thermostat method, in other mesoscale simulations \cite{Gross2010b,Gross2011}.

\section{acknowledgments}
Nima H. Siboni gratefully acknowledges the financial support from the Deutsche Forschungsgemeinschaft (German Research Foundation) through grant GSC 111. ICAMS gratefully acknowledges funding from its industrial sponsors, the state of North-Rhine Westphalia and the European Commission in the framework of the European Regional Development Fund (ERDF).

\bibliographystyle{aipnum4-1.bst}
\bibliography{biblio-v2}
\end{document}